\documentclass[apj]{emulateapj}

\newcommand{\kms}{\ifmmode {\rm km\,s}^{-1} \else km\,s$^{-1}$\fi}
\newcommand{\ergs}{\ifmmode {\rm erg\,s}^{-1} \else erg\,s$^{-1}$\fi}
\newcommand{\lsim}{\stackrel{\scriptscriptstyle <}{\scriptstyle {}_\sim}}
\newcommand{\gsim}{\stackrel{\scriptscriptstyle >}{\scriptstyle {}_\sim}}
\newcommand{\et}{\mbox{et~al.}\ }
\newcommand{\eg}{\mbox{e.g.,}\ }
\newcommand{\ie}{\mbox{i.e.,}\ }
\newcommand{\hb}{H$\beta$}
\newcommand{\Hbeta}{\ifmmode {\rm H}\beta \else H$\beta$\fi}

\newcommand{\civ}{C\,{\sc iv}} 

\newcommand{\mgii}{Mg\,{\sc ii}}

\newcommand{\Msigma}{\ifmmode M_{\rm BH} - \sigma \else $M_{\rm BH} - \sigma$\fi}
\newcommand{\Mbh}{\ifmmode M_{\rm BH} \else $M_{\rm BH}$\fi}
\newcommand{\mbh}{\ifmmode M_{\rm BH} \else $M_{\rm BH}$\fi}

\newcommand{\Msol}{\ifmmode M_{\odot} \else $M_{\odot}$\fi}

\slugcomment{Accepted by Astrophysical Journal Letters}

\shorttitle{Mass Functions of Distant Active Black Holes}
\shortauthors{Vestergaard et al.}

\begin{document}

\vspace{-2.5cm}

\title{Mass Functions of the Active Black Holes in Distant \\
   Quasars from the Sloan Digital Sky Survey Data Release 3
 }

\author{M. Vestergaard\altaffilmark{1,2}
X. Fan\altaffilmark{2}, C. A. Tremonti\altaffilmark{2,3},
Patrick S. Osmer\altaffilmark{4},
Gordon T. Richards\altaffilmark{5}}
\email{m.vestergaard@tufts.edu}

\altaffiltext{1}{Dept. of Physics and Astronomy,
Robinson hall, Tufts University, Medford, MA 02155 (present address)}
\altaffiltext{2}{Steward Observatory and Department of Astronomy,
933 N Cherry Avenue, Tucson, AZ 85721}
\altaffiltext{3}{Hubble Fellow}
\altaffiltext{4}{Graduate School, The Ohio State University,
230 N. Oval Mall, Columbus, OH, 43210}
\altaffiltext{5}{Department of Physics, Drexel University, 3141 Chestnut Street,
Philadelphia, PA 19104}



\begin{abstract}
We present the mass functions of actively accreting supermassive black holes 
over the redshift range 0.3 $\leq z \leq$ 5 
for a well-defined, homogeneous sample of 15,180 quasars from
the Sloan Digital Sky Survey Data Release 3 (SDSS DR3) within an 
effective area of 1644 deg$^2$. 
This sample is the most uniform statistically significant subset available
for the DR3 quasar sample. It was used for the DR3 quasar 
luminosity function, presented by Richards et al., and is the only sample 
suitable for the determination of the SDSS quasar black hole mass function.
The sample extends from $i$ = 15 to $i$ = 19.1 at $z \lsim 3$ and to 
$i$ = 20.2 for $z \gsim 3$. 
The mass functions display a rise and fall in the space density distribution of
active black holes at all epochs. 
Within the uncertainties the high-mass decline is consistent with a constant slope
of $\beta \approx -3.3$ at all epochs. This slope is similar to the bright end slope
of the luminosity function for epochs below $z = 4$.
Our tests suggest that the down-turn toward lower mass values is due to incompleteness 
of the quasar sample with respect to black hole mass. 
Further details and analysis of these mass functions will be
presented in forthcoming papers.  
\end{abstract}


\section{Introduction}

We live in exciting times. Since their discovery, the power-house of 
quasars was believed to be actively accreting supermassive black holes 
(\eg Salpeter 1964; Zeldovich \& Novikov 1964). Owing to recent advances, 
not only have we been able to confirm the existence of such dense, powerful 
objects (e.g., Ghez \et 2005; Genzel \et 2003; Harms \et 1994) the determination 
of their mass has become a common focus of many studies, in spite of the 
typically non-trivial nature of this process. The reasoning is quite simple:
(a) for local galaxies, the early studies revealed a close connection between 
the black hole mass and the mass, luminosity, and velocity dispersion of the 
spheroidal component of the host galaxy (\eg Magorrian \et 1998; Ferrarese \& 
Merritt 2000; Gebhardt \et 2000a, 2000b) suggesting an intimate connection 
between their evolutions and (b) recent semi-analytical and hydrodynamical 
simulations (\eg Granato \et 2004; Springel \et 2005; Cattaneo \et 2006; Kang \et 2006) 
show that black hole evolution and activity may have an important influence on the
properties of massive elliptical galaxies. Mapping the demography and mass
of the black holes in the centers of galaxies across the history of the Universe
is a first step to understanding
this connection. 
Since a fraction of the matter accreted onto black holes as they grow is 
converted into radiation, the cosmic evolution of the quasar luminosity 
function helps constrain the accretion and growth history of black holes 
and can therefore also help us understand the connection with galaxy evolution.
Yet certain degeneracies limit what we can learn from the luminosity
functions (\eg Wyithe \& Padmanabhan 2006) owing to the unknown degree of the 
radiative efficiency and the mass accretion rate and how these parameters 
evolve and depend on black hole properties.

In principle, these degeneracies can be broken by combining the luminosity
and mass functions.  Recently, Richards \et (2006, hereafter R06)
presented the luminosity function and its cosmic evolution from redshift 5 
to redshift 0.3 for a well-defined, homogeneous subset of 15,180 quasars from 
the Sloan Digital Sky Survey (SDSS) Data Release 3 (DR3) quasar catalog of over 
46,400 quasars for which the quasar selection function is well-determined.
Therefore, the determination of the mass function for this specific quasar 
sample is of high interest.
In this Letter we present this mass function of actively accreting black 
holes and its cosmic evolution from redshift 5 to 0.3.

To determine the mass of the black holes we adopt the scaling relations
based on the broad emission line widths and nuclear continuum luminosities
(\eg Vestergaard 2002; Warner \et 2003; Vestergaard \& Peterson 2006)
because (a) they are anchored in robust black hole mass determinations of
active black holes in the nearby Universe (Peterson \et 2004; Onken \et 2004), 
(b) several lines of evidence exist in favor of our application of this method 
to the most distant active black holes (see Vestergaard 2007 for details), and 
(c) the associated uncertainties on the mass estimates are among the lowest for 
the available mass estimation methods for distant sources. 
A cosmology of H$_0$ = 70 ${\rm km~ s^{-1} Mpc^{-1}}$,
$\Omega_{\Lambda}$ = 0.7, and $\Omega_m$ = 0.3 is used throughout.

\section{Spectral Measurements }\label{data}

As noted above, we use the quasar sample presented by R06 because
it is well-defined and has a well-understood selection function;
this is not the case for the full DR3 or the DR5 sample. Therefore it is 
currently also the only sample suitable for the determination of 
the black hole mass function.  The reader is referred to R06 for 
details on the sample and its selection.

In order to estimate the mass of the central
black hole in each quasar we need to measure the widths of each 
of the \hb, \mgii, and the \civ{} emission lines and the monochromatic
nuclear continuum luminosity near these emission lines. 
Since quasar spectra contain contributions from other emission 
components which often `contaminate' the line and continuum components 
of interest, we model each of these emission contributions so to 
obtain reliable measurements following a decomposition of the modeled 
components.
In the following we provide only a very brief summary of our data handling.
A more detailed account of our processing of the SDSS spectra will be
presented  by M.\ Vestergaard et al., (in preparation).

Each spectrum is corrected for Galactic reddening based on the
E($B - V$) value relevant for the quasar (Schneider \et 2005)
and the Galactic extinction maps by Schlegel \et (1998).
The continuum components were modeled using:
(a) a nuclear power-law continuum, (b) an optical-UV iron
line spectrum (Veron \et 2004; Vestergaard \& Wilkes 2001), 
(c) a Balmer continuum, and (d) a host galaxy spectrum (for objects 
at $z \leq$ 0.5 for which the galaxy contribution is strongest 
and is best characterized; Bruzual \& Charlot 2003).  

The continuum components were subtracted and the emission 
lines were then modeled with multiple Gaussian functions so to obtain 
smooth representations of the data. This allows us to eliminate most 
narrow absorption lines, noise spikes, and the contribution from the 
Narrow Emission Line Region, which is by far the strongest in the optical region. 
A single Gaussian component, not to exceed a FWHM of 600 \kms{}, was
used to model and subtract the narrow line components.
The line widths of \hb, \mgii, and \civ{} were measured for all
emission lines with model line peaks above
three times the median flux density error across the emission line
(\ie $\geq 3\sigma$ peaks).

All \mgii{} and \civ{} profiles with strong absorption were discarded 
from further analysis.  They were identified by visual inspection 
of the spectra of the quasars listed in the catalog of Trump \et (2006) to 
have absorption and of the quasars with redshifts between 1.4, when
\civ{} enters the observing window, and 1.7.  These quasars are prone to 
have strong \civ{} absorption but this redshift range is not covered by 
Trump \et (2006).

Of the 15,180 quasars on which the DR3 quasar luminosity function (R06) is based,
black hole mass estimates were possible for 14,434 quasars (95\%).

\begin{figure}
\epsscale{0.75}
\plotone{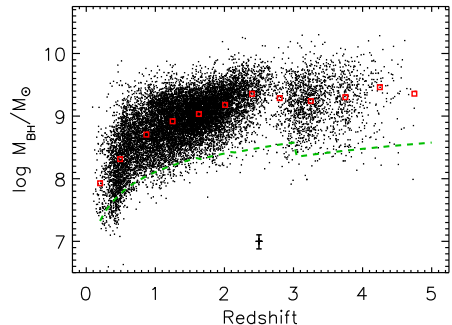}
\caption{Redshift distribution of the black hole masses of our quasar sample
selected from SDSS DR3. Of the 15,343 quasars detected in the selected sky area
at $z \geq 0.2$, mass estimates were possible for 14,584 quasars.
Of these, 14,434 quasars have redshifts between 0.3 and 5.0.  The median mass in 
each redshift bin shown in Figure~\ref{mfm.fig} is marked with the (red) open box. 
The median propagated black hole mass measurement error is shown in the lower 
part of the diagram.
For reference, the (green) dashed curve indicates the faint SDSS flux \mbox{limits folded 
with a line width of 2000 \kms.}
\label{Mz.fig}}
\end{figure}

\section{Black Hole Mass Estimates} \label{mbh}

The specific mass scaling relations we use are
equations 5 and 7 of Vestergaard \& Peterson (2006) based on the FWHM of \hb\ and 
\civ{} and a new relationship for the \mgii\ emission line based on high-quality 
data from the SDSS DR3 quasar sample (M.\ Vestergaard et al., in preparation).
For each quasar a black hole mass estimate is computed for each of the
\hb, \mgii, and \civ{} emission lines rendered 
suitable for mass estimates (\S~\ref{data}). 
The final black hole mass was computed as the variance weighted 
average of the individual mass estimates. The mass estimates based on 
the individual emission lines match well within their errors.

Figure~\ref{Mz.fig} shows the redshift distribution of the black hole masses.
Its detailed shape is mainly a consequence of the survey properties and limitations: 
(a) the scarcity of quasars at 2.2 $\leq z \leq$ 3.0 is due to 
inefficient quasar selection, discussed later and (b) the lower boundary is 
determined mainly by the faint survey flux limits ($i <$ 19.1 mag at $z <$ 3.0; 
$i <$ 20.2 mag at $z \geq 3$). While the bright flux limit ($i >$ 15 mag) of SDSS 
is expected to eliminate the most massive black holes at $z <$ 1, this does not
have a strong impact on the current study. 
Judging from the comparison made (Jester \et 2005) between the Bright Quasar Survey 
(hereafter BQS, Schmidt and Green 1983) and SDSS the surface density of quasars brighter than
$i$ = 15.0 is such that at least 6 such bright quasars are expected in the 1622 
square degrees of the current sample; this is a lower limit since the BQS itself
is incomplete. If these bright quasars follow the same redshift distribution as 
the BQS quasars detected by SDSS, they will mostly lie at $z \lsim$ 0.25, below
the minimum redshift bound (0.3) of the mass functions.
Also, no upper 
limit in line width is imposed that would place an upper bound on the black hole 
masses. While we do deselect low quality emission line profiles (see \S~\ref{data}) 
this process applies to all profile widths (and thus all black hole masses). 
Given our large sample size (nearly 15,000 quasars) and that we know of no 
selection effects that would systematically deselect the most massive black holes, 
it is fair to conclude that the upper bound in black hole mass is real.
The mass distribution in Figure~\ref{Mz.fig} therefore confirms with larger 
numbers the results from previous studies (\eg Corbett et al. 2003; Warner \et 2003; 
Netzer 2003; McLure \& Dunlop 2004; Vestergaard 2004; 
Kollmeier \et 2006; Netzer \& Trakhtenbroot 2007; see also Vestergaard 2006, 2007; Shen \et 2007) that 
find the detected quasars at high redshift have black hole masses between a few 
times 10$^8$\Msol{} to 10$^{10}$\Msol{} with typical mass of order 10$^9$ \Msol.

\section{Black Hole Mass Functions}

\begin{figure}
\includegraphics[angle=90,scale=0.38]{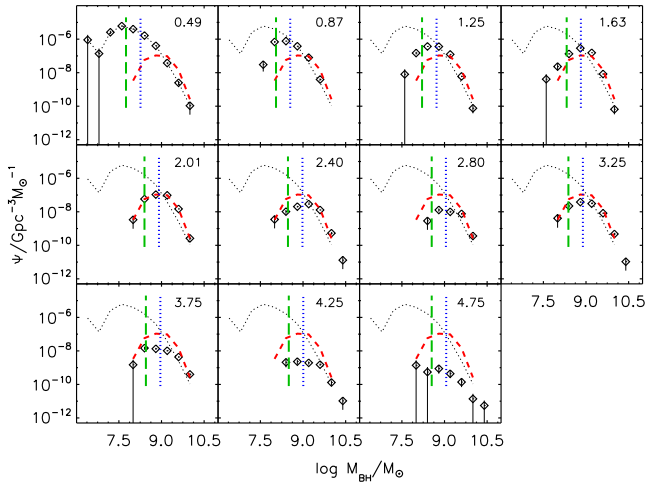}
\caption{Mass functions of active supermassive black holes in the quasar
sample at a range of redshifts (open diamonds; error bars denote the Poisson
statistical uncertainty). The mean redshift of each redshift bin is labeled. 
The boundaries of the redshift bins are: 0.3, 0.68, 1.06, 1.44, 1.82, 2.2, 
2.6, 3.0, 3.5, 4.0, 4.5, 5.0. In each panel the $z$ = 0.49 and 2.01 mass 
functions are also shown as (black) dotted and (red) dashed curves for reference. 
The vertical (green) dashed and (blue) dotted lines mark the masses of two 
fiducial black holes at the mean redshift of the z-bin, emitting a luminosity 
equal to the survey flux limit, and around which the broad emission line gas is
moving at 2000 \kms{} and 3500 \kms{} (FWHM), respectively. 
Incompleteness is expected for black hole masses below the 2000 \kms{} equivalent
fiducial mass value, but may also affect bins of higher mass values.
See text for details and discussion.
\label{mfm.fig}}
\end{figure}

The differential quasar black hole mass function, $\Psi$(\mbh,$z$) is 
the space density of quasar black holes (\ie the number of black holes 
per unit comoving volume) per unit black hole mass as a function of
black hole mass and redshift. We calculate this space density 
in different mass and redshift bins using the 1/$V_a$ method
following Warren, Hewett, and Osmer (1994), 
where $V_a$ is the accessible volume, defined by Avni and Bahcall (1980). 
We adopt the method of Fan \et (2001) of including the selection
function of the quasars: we interpolate the selection function between 
the grid points defined by the quasar luminosity and redshift and 
integrate over the selection function in determining $V_a$; see 
equation (6) of Fan et al. 
The SDSS DR3 quasar selection function for our quasar sample is 
published by R06. 
The mass function (and its statistical uncertainty) is given by
\begin{mathletters}
\begin{eqnarray}
\Psi (<\!M\!>,<\!z\!>) = \sum_i \frac{1}{V_{a,i} \; \Delta M_{\rm BH}}, \\
\sigma(\Psi) = \left[ \sum_i \left(\frac{1}{V_{a,i} \; \Delta M_{\rm BH}}\right)^2 \right]^{1/2},
\end{eqnarray}
\end{mathletters}
\noindent
where $<\!M\!>$ and $<\!z\!>$ are the average mass and redshift of each 
mass bin of size $\Delta M_{\rm BH}$ and redshift bin of size $\Delta z$ 
and index $i$ refers to each individual quasar in this mass and redshift 
bin.  
In Figure~\ref{mfm.fig} we show the mass functions at a range of redshifts.
For better comparison, the redshift bins are chosen to be those used 
for the luminosity function for this sample (R06). 
The data points with large errors are typically due to very few quasars 
residing in those mass bins. 

\section{Discussion and Summary}

For each redshift bin in Figure~\ref{mfm.fig} the mass function
displays a rise and fall in the mass distribution.  The decline 
toward higher masses is expected 
to be intrinsic to the quasar population (\S~\ref{mbh}); while the 
uncertainty in the mass estimates will affect the accuracy of the 
slope value, it is too small to be responsible for the decline.
We determine\footnote{The slope of the mass functions at $z < 2$ 
is determined for masses above 9.2\,dex 
and at $z > 2$ for \mbh{} $\geq$ 9.6\,dex,
since these mass bins are deemed (essentially) completely populated.
To account somewhat for the uncertainties in the \mbh{} values 
and that the mass bin population varies, the uncertainty in 
a given mass bin was estimated as the propagated uncertainty in the 
absolute mass estimates (\eg Table~5 of Vestergaard \& Peterson 2006)
given the Poisson population statistics.
} a typical high-end slope $\beta$ of about $-$3.3 
($\Psi \propto M^{\beta}$) with uncertainties between 0.4 to 1.3
for the mass functions below a redshift of 4. 
At the mean redshifts of 4.25 and 4.75 the high-end slope `flattens'
to $\beta \approx -2.9 \pm 1.1$ and $\beta \approx -1.9 \pm 1.7$,
respectively.  The mass function at redshift 
4.75 is based on very few black holes at each mass bin, especially 
for \mbh{} $\geq 10^{10}$\Msol, so the shape is somewhat uncertain. 
The mass functions are clearly consistent with a constant slope across 
all epochs.  This is also easily verified in Figure~\ref{mfm.fig} 
by the reference curves for the mass functions at $z$=0.49 (black 
dotted) and at $z$= 2.01 (red dashed)  which also show the space 
density of high mass black holes slightly increases at $z \gsim$ 1.

The slope of the mass functions is consistent with the bright end 
slope of the quasar luminosity function ($\beta \approx -3.3$; Croom \et 
2004; R06) to within the uncertainties with the exception that the latter 
flattens above a redshift of 4 (to $\beta = - 2.5$; Schmidt \et 1995;
Fan \et 2001). Another difference 
is the relative normalizations. The mass functions exhibit a change in
space density for a given mass of only a factor of a few between $z$ = 0.5 
and $z$ = 2 (cf.\ the dotted and dashed curves in Fig~\ref{mfm.fig}), while 
the space density 
of quasars brighter than $\sim$\,$-$27 mag drops 2 orders of magnitude or more 
between redshifts 2 and 0.5 (Fig.~18, R06).  While there is no one-to-one 
correspondence between a given mass value and a given luminosity value 
since black holes radiate at a range of (Eddington) rates, the different
space density changes show that the cumulative volume density of $L/M$ 
($ \propto L/L_{\rm Edd}$) decreases toward low redshift. 
This confirms what we already know based on the declining space density of
active nuclei at later epochs (\eg R06): low redshift black holes are typically
less active.

In Figure~\ref{mfm.fig}, a characteristic peak shift to higher mass values 
with redshift is seen. It is important to establish 
the reality of these peaks and their shifts,
since this would have intriguing cosmological implications. 
For one, it would be a clear manifestation of cosmic down-sizing: the most massive 
black holes were more actively accreting early in the universe and with time
progressively less massive black holes dominate the population of 
actively accreting black holes.
Alternatively, the down-turn toward lower mass values may simply be due 
to incompleteness of the sample, as the source selection is based on the 
source luminosity and broad-band colors, not on black hole mass. 
As a result, we will not obtain a sharp lower boundary in mass (as we do
in the luminosity distribution) but a decrease in space density below the survey flux limits,
owing to the distribution of black hole masses at a given luminosity.
Unfortunately, the SDSS does not go deep enough for us to make meaningful 
cuts in black hole mass for the current sample, as will be clear later.

The best way to assess which part of the down-turn is intrinsic requires 
detailed simulations that include the uncertainties in the mass estimates, 
the selection function, and the flux limits. Since this is non-trivial we 
will address this in a forthcoming paper (B.\ Kelly et al., in preparation). 
Here, we instead perform crude checks as follows. Firstly, in Figure~\ref{mfm.fig} 
we mark the masses of a fiducial source at the mean bin redshift 
with a continuum luminosity equivalent to the flux limit at that 
redshift and a broad line width of 2000 \kms{} (green dashed vertical line)
and 3500 \kms{} (blue dotted vertical line), respectively; 2000 \kms{} is
the canonical lower line width for which a source is a bona-fide quasar and
3500 \kms{} is the median linewidth for nearby quasars. 
There are objects below the (green) dashed line since the observed $i$-band 
magnitude contains emission contribution in addition to the power-law continuum
luminosity used for the mass estimates (see \S~\ref{data}).
Nonetheless, these fiducial masses do yield a guideline location of the
SDSS flux limits in the mass parameter space, and their location
close to the peak of the mass distribution suggests that the peak shift
is simply due to the survey flux limits.

Given the non-trivial relationship between the $i$-band magnitude and the
monochromatic luminosity, we also performed a simple simulation with the
aim of minimizing selection effects.
We generated a mock catalog of quasars covering redshifts between 0.2 to 5.0
and bolometric luminosities between $10^{42}$ \ergs{} and $10^{48}$ \ergs{}.
For each ($L,z$) grid point we computed the distribution of black hole masses
by applying the typical {\em observed} line width distribution for these
$L$ and $z$ values.  The mass functions based on this mock catalog (not shown)
do not display the narrowly peaked distributions seen in Figure~\ref{mfm.fig},
but display a slower continued rise in the distribution below the high mass 
fast decline and a sharp downturn at the lowest masses which coincides with 
the lower luminosity bound.  No downturn is seen at the low mass end if a 
sharp cut in black hole mass is imposed. 
Hence, this mock catalog confirms the earlier indication that the low-mass
downturn in the observed mass functions is a consequence of the relatively 
narrow luminosity range
probed by SDSS broadened by the line width distribution, especially at 
higher redshift. We thus conclude that the peaks and their shifts with redshift 
are mainly due to incompleteness of active black holes near the survey flux limits.
These crude checks are, however, unable to verify which part of the turnover
is trustworthy. The mass functions in the lower redshift bins display a
slower turnover, part of which may well be real. B.\ Kelly et al.,
(in preparation) will present more advanced simulations that address these 
and related issues.

The space density decline at $z$\,$\approx$\,2.8 relative to $z$\,=\,2.4 and $z$\,=\,3.25
in Figure~\ref{mfm.fig} is caused by a much lower selection efficiency 
(by a factor of 8; Fig.~6 of R06) as quasars
at these epochs coincide with the stellar loci in the selected color-spaces.
This suggests 
that the completeness level is slightly overestimated at these epochs.
This situation is unfortunate since the space density of quasars is
known to peak between redshifts of 2 and 3 (\eg Osmer 1982; 
Warren \et 1994; Schmidt, Schneider, \& Gunn 1995; 
Fan \et 2001).  It is thus not possible 
with the current sample to get a complete picture of how the space density 
distribution of black holes behave during this important epoch. 
The Large Bright Quasar Survey (hereafter LBQS; Hewett \et 1995) of about 
1000 quasars, extending to a redshift of 3, 
offers an opportunity to address these issues further, since this sample
is not selected based on broad-band color. The LBQS quasar mass function
will be presented by Vestergaard \& Osmer (in preparation) and
discussed in concert with the DR3 mass function by M.\ Vestergaard et al.\
(in preparation), where the mass functions will also be compared to previous 
work.

\acknowledgments

We gratefully acknowledge financial support from NSF grant AST 03-07384 (XF, MV),
HST grant HST-AR-10691 (MV) and Hubble Fellowship grant HST-HF-01192.01 (CAT) awarded
by the Space Telescope Science Institute, which is operated by the Association of
Universities for Research in Astronomy, Inc., for NASA, under contract NAS5-26555,
a Packard Fellowship for Science and Engineering (XF), and Sloan Research
Fellowships (XF, GTR).
Funding for the SDSS has been provided by
the Alfred P. Sloan Foundation, the Participating Institutions, the
National Aeronautics and Space Administration, the National Science
Foundation, the U.S. Department of Energy, the Japanese
Monbukagakusho, and the Max Planck Society. The SDSS Web site is
http://www.sdss.org/.



{\it Facilities:} \facility{SDSS}



\begin{thebibliography}{}
\bibitem[Avni \& Bahcall(1980)]{1980ApJ...235..694A} Avni, Y., \& Bahcall, J.~N.\ 1980, \apj, 235, 694
\bibitem[]{bc03} Bruzual, X. \& Charlot, X. 2003, \mnras, 344, 1000
\bibitem[Cattaneo et al.(2006)]{2006MNRAS.370.1651C} Cattaneo, A., Dekel, 
	A., Devriendt, J., Guiderdoni, B., \& Blaizot, J.\ 2006, \mnras, 370, 1651
\bibitem[]{} Corbett, E.A., et al.\ 2003, \mnras, 343, 705
\bibitem[Croom et al.(2004)]{540} Croom, S. M., Smith, R. J., Boyle, B.J., Shanks, T., Miller, L.,
        Outram, P.J., and Loaring, N. S., 2004,  \mnras, 349, 1397
\bibitem[]{542} Fan, X. \et 2001, \aj, 121, 54  
\bibitem[Ferrarese and Merritt 2000]{543} Ferrarese, L. \& Merritt, D. 2000, \apj, 539, L9
\bibitem[Gebhardt et al. 2000a]{2000ApJ...539L..13G} Gebhardt, K. et al. 2000a, \apj, 539, L13
\bibitem[Gebhardt et al. 2000b]{2000ApJ...543L...5G} Gebhardt, K. et al. 2000b, \apj, 543, L5
\bibitem[Genzel et al.(2003)]{2003ApJ...594..812G} Genzel, R., et al.\ 2003, \apj, 594, 812
\bibitem[Ghez et al.(2005)]{2005ApJ...620..744G} Ghez, A.~M., Salim, S., 
  Hornstein, S.~D., Tanner, A., Lu, J.~R., Morris, M., Becklin, E.~E., \& 
  Duch{\^e}ne, G.\ 2005, \apj, 620, 744 
\bibitem[Granato et al.(2004)]{2004ApJ...600..580G} Granato, G.~L., De 
	Zotti, G., Silva, L., Bressan, A., \& Danese, L.\ 2004, \apj, 600, 580
\bibitem[Harms et al.(1994)]{1994ApJ...435L..35H} Harms, R.~J., et al.\ 1994, \apjl, 435, L35
\bibitem[]{553} Hewett, P. C., Foltz, C. B., \& Chaffee, F.H. 1995, \aj, 109, 1498 
\bibitem[Jester et al.(2005)]{2005AJ....130..873J} Jester, S., et al.\ 2005, \aj, 130, 873
\bibitem[Kang et al.(2006)]{2006ApJ...648..820K} Kang, X., Jing, Y.~P., \& Silk, J.\ 2006, \apj, 648, 820
\bibitem[Kollmeier et al.(2006)]{2006ApJ...648..128K} Kollmeier, J.~A., et al.\ 2006, \apj, 648, 128
\bibitem[Magorrian et al. 1998]{1998AJ....115.2285M} Magorrian, J. et al. 1998, \aj, 115, 2285
\bibitem[]{} Mclure, R. \& Dunlop, J. 2004, \mnras, 352, 1390
\bibitem[Netzer 2003]{558} Netzer, H. 2003, \apj,   583, L5
\bibitem[]{} Netzer, H. \& Traktenbroot, B. 2007, \apj,   654,754 
\bibitem[Onken et al. 2004]{2004ApJ...615..645O} Onken, C.A., et al. 2004, \apj,   615, 645
\bibitem[Osmer(1982)]{1982ApJ...253...28O} Osmer, P.~S.\ 1982, \apj, 253, 28
\bibitem[Peterson et al. 2004]{2004ApJ...613..682P} Peterson, B.M. et al. 2004, \apj,   613, 682
\bibitem[Richards et al.(2006)]{2006AJ....131.2766R} Richards, G.~T., et 
	al.\ 2006, \aj, 131, 2766 (R06)
\bibitem[Salpeter(1964)]{1964ApJ...140..796S} Salpeter, E.~E.\ 1964, \apj, 140, 796
\bibitem[Schlegel et al.(1998)]{1998ApJ...500..525S} Schlegel, D.~J., 
	Finkbeiner, D.~P., \& Davis, M.\ 1998, \apj, 500, 525 
\bibitem[Schneider et al. 2005]{567} Schneider, D.\ et al. 2005, \aj,  130, 367
\bibitem[Schmidt \& Green(1983)]{1983ApJ...269..352S} Schmidt, M., \& 
	Green, R.~F.\ 1983, \apj, 269, 352
\bibitem[Schmidt, Schneider, \& Gunn (1995)]{1995AJ....110.68S} Schmidt, M., Schneider, D.~P.,
        \& Gunn, J. E. 1995, \aj, 110, 68
\bibitem[]{572} Shen, Y., Greene, J.\ E., Strauss, M., Richards, G.\ T., Schneider, D.P.\ 2007, preprint (arXiv:0709.3098v1)
\bibitem[Springel et al.(2005)]{2005ApJ...620L..79S} Springel, V., Di 
  	Matteo, T., \& Hernquist, L.\ 2005, \apjl, 620, L79
\bibitem[Trump et al.(2006)]{2006ApJS..165....1T} Trump, J.~R., et al.\ 
	2006, \apjs, 165, 1  
\bibitem[V{\'e}ron-Cetty et al.(2004)]{2004A&A...417..515V} 
	V{\'e}ron-Cetty, M.-P., Joly, M., \& V{\'e}ron, P.\ 2004, \aap, 417, 515 
\bibitem[Vestergaard(2002)]{2002ApJ...571..733V} Vestergaard, M.\ 2002, \apj, 571, 733
\bibitem[Vestergaard(2004)]{2004ApJ...601..676V} Vestergaard, M.\ 2004, \apj, 601, 676
\bibitem[Vestergaard(2006)]{2006NewAR..50..817V} Vestergaard, M.\ 2006, New Astronomy Review, 50, 817
\bibitem[Vestergaard(2007)]{582} Vestergaard, M.\ 2007, in the STScI Spring Symposium 
	2007 on 'Black Holes', ed. A. Koekemoer, Cambridge University Press, in press
\bibitem[Vestergaard \& Peterson(2006)]{2006ApJ...641..689V} Vestergaard, 
	M., \& Peterson, B.~M.\ 2006, \apj, 641, 689
\bibitem[Vestergaard \& Wilkes(2001)]{2001ApJS..134....1V} Vestergaard, M., \& 
	Wilkes, B.~J.\ 2001, \apjs, 134, 1
\bibitem[Warner et al.(2003)]{2003ApJ...596...72W} Warner, C., Hamann, F., 
	\& Dietrich, M.\ 2003, \apj, 596, 72
\bibitem[Warren et al.(1994)]{1994ApJ...421..412W} Warren, S.~J., Hewett, P.~C., 
	\& Osmer, P.~S.\ 1994, \apj, 421, 412
\bibitem[Wyithe \& Padmanabhan(2006)]{2006MNRAS.372.1681W} Wyithe, 
J.~S.~B., \& Padmanabhan, T.\ 2006, \mnras, 372, 1681
\bibitem[]{594} Zeldovich, Y.\ B.\ \& Novikov, I. 1964, Sov.\ Phys.\ Dokl.\ 158, 811
\vspace{-0.4cm}
\end{thebibliography}
\end{document}